\documentclass[12pt]{article}

\usepackage{graphicx}
\usepackage{lscape}
\usepackage{epsfig}
\usepackage{citesort}
\usepackage{amssymb}
\usepackage{amsmath}
\usepackage{multirow}

\def\qq{\langle \bar q q \rangle}

\newcommand{\be}{\begin{equation}}
\newcommand{\ee}{\end{equation}}
\newcommand{\bea}{\begin{eqnarray}}
\newcommand{\eea}{\end{eqnarray}}

\def\R1{\varepsilon_1}
\def\E8{\varepsilon_8}

\def\s1{\hat s}
\def\ds{\displaystyle}

\newcommand{\bd}{\begin{displaymath}}
\newcommand{\ed}{\end{displaymath}}

\def\R1{\varepsilon_1}
\def\E8{\varepsilon_8}

\def\ds{\displaystyle}
\def\beq{\begin{equation}}
\def\eeq{\end{equation}}
\def\bea{\begin{eqnarray}}
\def\eea{\end{eqnarray}}
\def\beeq{\begin{eqnarray}}
\def\eeeq{\end{eqnarray}}

\def\nnb{\nonumber}

\def\lla{\left<}
\def\rra{\right>}

\def\nnb{\nonumber}
\def\la{\langle}
\def\ra{\rangle}
\def\ba{\begin{array}}
\def\ea{\end{array}}

\def\xis0{{\Xi^{*0}}}

\def\qq{\la \bar q q \ra}

\def\g5{\gamma_5}

\def\es{\!\!\! &=& \!\!\!}

\begin{document}
\title{
         {\Large
    {\bf Doubly heavy spin--1/2 baryon spectrum in QCD}
         }
      }
\author{\vspace{1cm}\\
{\small  T. M. Aliev$^1$ \thanks {e-mail: taliev@metu.edu.tr}}\,\,
{\small  K. Azizi$^2$ \thanks {e-mail: kazizi@dogus.edu.tr}}\,\,,
{\small M. Savc{\i}$^1$ \thanks
{e-mail: savci@metu.edu.tr}} \\
{\small $^1$ Department of Physics, Middle East Technical University,
06531 Ankara, Turkey}\\
{\small $^2$ Department of Physics, Do\u gu\c s University,
Ac{\i}badem-Kad{\i}k\"oy, 34722 \.{I}stanbul, Turkey}}

\date{}

\begin{titlepage}
\maketitle
\thispagestyle{empty}
\begin{abstract}
We calculate the mass and residue of the heavy spin--1/2 baryons containing 
two heavy b or c quarks in the framework of QCD sum rules. We use the most
general form of the interpolating current in its symmetric and anti-symmetric
forms with respect to the exchange of  heavy quarks, to calculate the
two-point correlation functions describing the baryons under consideration.
A comparison of the obtained results with existing predictions from various
approaches is also made.
\end{abstract}
~~~PACS number(s): 11.55.Hx,  14.20.-c, 14.20.Mr, 14.20.Lq
\end{titlepage}

\section{Introduction}

The quark model practically describes all phenomena in hadron physics very
successfully. This model predicts the existence of hadrons containing two heavy
quarks  and one can estimate their masses in this framework \cite{Rbwtq01}.
The first experimental observation of the doubly heavy baryons was reported by
SELEX Collaboration \cite{Rbwtq02}, where the double  charmmed $\Xi^{+}_{cc}$ and
$\Xi^{++}_{cc}$ baryons were found. This observation was confirmed by
measurement of a different weak decay mode \cite{Rbwtq03,Rbwtq04}.
The experimental discovery of the doubly heavy baryons and study of their
properties constitutes one of the main directions of the physics program at LHC.
 (for a review on doubly heavy
baryons, see for example \cite{Rbwtq05}).

In this prospect, it would be interesting to present reliable
theoretical predictions on properties of these baryons. The masses of
the doubly heavy baryons have been estimated in the frameworks of the quark 
\cite{Rbwtq06,Rbwtq07,Rbwtq08}  and MIT bag  \cite{Rbwtq09} models. In order to
study the properties of these hadrons in a model independent way, the
QCD sum rules method \cite{Rbwtq10} based on QCD Lagrangian is
one of the most reliable approaches. In the present work, we calculate
the masses and residues of the doubly heavy spin--1/2 baryons within QCD
sum rules formalism. We consider both symmetric and anti-symmetric
currents with respect to the exchange of  heavy quarks defining the
baryons under consideration in their most general forms. It should be
noted that the masses of doubly heavy baryons with spin--1/2 have been
calculated within the same framework using the Ioffe current in
\cite{Rbwtq11,Rbwtq12} and general current in \cite{Rbwtq13}. However, the
obtained expressions for mass sum rules  in  \cite{Rbwtq13} at Ioffe
current case does not exactly  reduce to those presented in
\cite{Rbwtq11,Rbwtq12} for Ioffe current. The masses of the doubly heavy
baryons with spin--3/2 have also been investigated in \cite{Rbwtq13}
and \cite{Rbwtq14}.

The plan of this work is as follows. In next section, the QCD sum rules for
the masses and residues of the doubly spin--1/2 baryons are obtained. Section 3
is devoted to the numerical analysis of the sum rules for physical 
quantities under consideration. This section   encompasses also our
comparison of the obtained results with those existing in the
literature as well as our concluding remarks.

\section{Sum rules for the masses and residues of the doubly heavy baryons
with spin--1/2}

Before presenting the detailed calculation of sum rules for the masses and
residues of the doubly heavy baryons let us discuss the ground states of
these baryons in quark model. For the ground state of doubly  heavy
baryons with the identical heavy quarks, $\Xi^{(*)}_{QQ}$ and
$\Omega^{(*)}_{QQ}$, the pair of heavy quarks form a diquark  with total
spin of 1. Adding then the spin--1/2 of light quark, we have two states with
total spin--1/2 or spin--3/2. The baryons with (without) star stand for
spin 3/2 (1/2) doubly heavy baryons during the text. For these states, the interpolating current
should be symmetric with respect to the exchange of the  heavy
quarks fields. For the states containing two different heavy quarks, in addition
to the previous case, i.e., total spin of diquark equal to one, 
the diquark can also have the total spin zero, which leads to the total
spin 1/2 of these states. Obviously, the interpolating current of these states
(usually these states are denoted by $\Xi'_{bc}$  and $\Omega'_{bc}$) are
anti-symmetric with respect to two heavy quarks fields. In the
present work, we deal only with spin--1/2 doubly heavy baryons.

After these preliminary remarks, let us come back to our main problem, i.e.,
calculation of the masses and residues of the doubly heavy baryons.
The interpolating currents here play
the role of the wave functions in the quark model. The general expressions of the interpolating
currents for the spin--1/2 doubly heavy baryons in their  symmetric and
anti-symmetric forms  can be written as: 
\begin{eqnarray}
 \eta^{S}&=&\frac{1}{\sqrt{2}}\epsilon_{abc}\Bigg\{(Q^{aT}Cq^b)\gamma_{5}Q'^c+
(Q'^{aT}Cq^b)\gamma_{5}Q^c+\beta(Q^{aT}C\gamma_{5}q^b)Q'^c+\beta(Q'^{aT}C
\gamma_{5}q^b)Q^c\Bigg\},\nonumber\\
\eta^{A}&=&\frac{1}{\sqrt{6}}\epsilon_{abc}\Bigg\{2(Q^{aT}CQ'^b)\gamma_{5}q^c+
(Q^{aT}Cq^b)\gamma_{5}Q'^c-(Q'^{aT}Cq^b)\gamma_{5}Q^c+2\beta(Q^{aT}C
\gamma_{5}Q'^b)q^c\nonumber\\
&+&\beta(Q^{aT}C\gamma_{5}q^b)Q'^c-\beta(Q'^{aT}C\gamma_{5}q^b)Q^c\Bigg\},
\end{eqnarray}
where $\beta$ is an arbitrary auxiliary parameter.
The case,  $\beta=-1$ corresponds to the Ioffe current.
Here $C$ stands for the charge conjugation operator, $T$ denotes the transposition,
$a$, $b$, and $c$ are the color indices; and $Q$ and $q$ correspond to the heavy and
light quarks fields, respectively. The interpolating current with the light quark $u$ or $d$
corresponds to the $\Xi_{QQq}$, and with $s$ to the $\Omega_{QQq}$ baryons, respectively.
Here we would like to note once more that in the symmetric part, both heavy
quarks may be identical or different, but in the anti-symmetric part two
heavy quarks must be different.

For calculation of the masses and residues of the doubly heavy baryons within
the QCD sum rules, we consider the following correlation function:
\begin{eqnarray}
 \Pi^{S(A)}(q)=i \int d^4x e^{i q x} \langle 0|{\cal T}
\Big\{\eta^{S(A)}(x) \bar \eta^{S(A)}(0)\Big\}|0\rangle,
\end{eqnarray}
where $q$ is four-momentum of the doubly heavy baryon. This correlation
function can be written in terms of two independent structures
\begin{eqnarray}\label{str}
\Pi^{S(A)}(q)=\rlap/{ q}\Pi_1^{S(A)}(q^2)+U \Pi_2^{S(A)}(q^2),
\end{eqnarray}
where $U$ stands for unit matrix.

The above mentioned correlation function can be calculated in two different manners. From
one side, it is calculated in terms of hadronic parameters called physical
or phenomenological part. On the other side, it is evaluated in terms
of quarks and gluons and their interactions with each other and QCD vacuum
called theoretical or QCD side. Equating these two different representations
of the correlation function to each other under the  flag of the quark-hadron
duality assumption, we get sum rules for the physical quantities under
consideration. To suppress the contributions coming from the higher states
and continuum and obtain contribution of the ground state,
we apply the Borel transformation and continuum subtraction after performing
the Fourier integrals. These procedures bring two auxiliary parameters so
called Borel mass parameter and continuum threshold, which we shall
also find their working region to numerically analyze the obtained sum rules
in the next section.

Saturating the  correlation function   by a complete set of hadronic states with the same quantum numbers as the interpolating current and isolating the 
ground state baryons in
phenomenological side, we have
\begin{eqnarray}
 \Pi^{S(A)}(q)=\frac{\langle 0|\eta^{S(A)} (0)|B(q)\rangle \langle B(q)|
\bar \eta^{S(A)} (0)|0\rangle}{q^2-m_B^2}+...,
\end{eqnarray}
where dots refers to the  contribution of the higher states and continuum. The
matrix element $\langle 0|\eta (0)|B(q)\rangle$ for spin--1/2 baryons is
determined as
\begin{eqnarray}
 \langle 0|\eta (0)|B(q,s)\rangle=\lambda_B u(q,s),
\end{eqnarray}
where $\lambda_B$ is the residue and $u(q,s)$ is the Dirac spinor with $s$ stands
for the spin. Putting the above equations all together, and performing summation 
over spins, we get the following final representation for the phenomenological side:
\begin{eqnarray}
 \Pi^{S(A)}(q)=\frac{\lambda^{2}_{B^{S(A)}}(\rlap/{ q}+m_{B^{S(A)}})}{q^2-
m_{B^{S(A)}}^2}+...,
\end{eqnarray}
where we have only two independent Lorentz structures $\rlap/{ q}$ and $U$
discussed above.

In QCD side, the correlation function is calculated in deep Euclidean region 
with the help of operator product expansion (OPE). Using the Wick theorem and
performing all contractions of the quarks fields 
for the symmetric
 part, we get the following expression in terms of heavy and light quarks
propagators:
\begin{eqnarray}\label{tree expresion.m}
\Pi^{S}(q)&=&iA\epsilon_{abc}\epsilon_{a'b'c'}\int d^4x e^{i q x}\langle0\mid\Big\{-\gamma_{5}
S^{cb'}_{Q}S'^{ba'}_{q}S^{ac'}_{Q'}\gamma_{5}
-\gamma_{5}S^{cb'}_{Q'}S'^{ba'}_{q}S^{ac'}_{Q}\gamma_{5}
\nonumber\\&+&\gamma_{5}S^{cc'}_{Q'}\gamma_{5}Tr\Big[S^{ab'}_{Q}S'^{ba'}_{q}\Big]
+\gamma_{5}S^{cc'}_{Q}\gamma_{5}Tr\Big[S^{ab'}_{Q'}S'^{ba'}_{q}\Big]
\nonumber\\&+&\beta\Big( -\gamma_{5}S^{cb'}_{Q}\gamma_{5}S'^{ba'}_{q}S^{ac'}_{Q'}
-\gamma_{5}S^{cb'}_{Q'}\gamma_{5}S'^{ba'}_{q}S^{ac'}_{Q}-S^{cb'}_{Q}S'^{ba'}_{q}
\gamma_{5}S^{ac'}_{Q'}\gamma_{5}\nonumber\\&-&S^{cb'}_{Q'}S'^{ba'}_{q}\gamma_{5}
S^{ac'}_{Q}\gamma_{5}
+\gamma_{5}S^{cc'}_{Q'}Tr\Big[S^{ab'}_{Q}\gamma_{5}S'^{ba'}_{q}\Big]+S^{cc'}_{Q'}
\gamma_{5}Tr\Big[S^{ab'}_{Q}S'^{ba'}_{q}\gamma_{5}\Big]\nonumber\\&+&
\gamma_{5}S^{cc'}_{Q}Tr\Big[S^{ab'}_{Q'}\gamma_{5}S'^{ba'}_{q}\Big]+S^{cc'}_{Q}
\gamma_{5}Tr\Big[S^{ab'}_{Q'}S'^{ba'}_{q}\gamma_{5}\Big]\Big)\nonumber\\&+&
\beta^2\Big( -S^{cb'}_{Q}\gamma_{5}S'^{ba'}_{q}\gamma_{5}S^{ac'}_{Q'}-S^{cb'}_{Q'}
\gamma_{5}S'^{ba'}_{q}\gamma_{5}S^{ac'}_{Q}\nonumber\\&+&
S^{cc'}_{Q'}Tr\Big[S^{ba'}_{q}\gamma_{5}S'^{ab'}_{Q}\gamma_{5}\Big]+S^{cc'}_{Q}
Tr\Big[S^{ba'}_{q}\gamma_{5}S'^{ab'}_{Q'}\gamma_{5}\Big]
\Big)
\Big\}\mid 0\rangle,
\end{eqnarray}
where $S'=CS^TC$. In the case of
$Q\neq Q'$, the constant $A$ in the above equation takes the value $A=\frac{1}{2}$, 
while when $Q= Q'$  we have $A=1$ as a result of extra contractions between the
same quark fields. For the  anti-symmetric part we have
\begin{eqnarray}\label{tree expresion.m2}
 \Pi^{A}(q)&=&\frac{i}{6}\epsilon_{abc}\epsilon_{a'b'c'} \int d^4x e^{i q x}\langle0\mid
\Big\{2\gamma_{5}S^{cb'}_{Q}S'^{aa'}_{Q'}S^{bc'}_{q}\gamma_{5}
+\gamma_{5}S^{cb'}_{Q}S'^{ba'}_{q}S^{ac'}_{Q'}\gamma_{5}
\nonumber\\&-&2\gamma_{5}S^{ca'}_{Q'}S'^{ab'}_{Q}S^{bc'}_{q}\gamma_{5}
+\gamma_{5}S^{cb'}_{Q'}S'^{ba'}_{q}S^{ac'}_{Q}\gamma_{5}
\nonumber\\&-&2\gamma_{5}S^{ca'}_{q}S'^{ab'}_{Q}S^{bc'}_{Q'}\gamma_{5}
+2\gamma_{5}S^{ca'}_{q}S'^{bb'}_{Q'}S^{ac'}_{Q}\gamma_{5}
+4\gamma_{5}S^{cc'}_{q}\gamma_{5}Tr\Big[S^{ab'}_{Q}S'^{ba'}_{Q'}\Big]\nonumber\\
&+&\gamma_{5}S^{cc'}_{Q'}\gamma_{5}Tr\Big[S^{ab'}_{Q}S'^{ba'}_{q}\Big]
+\gamma_{5}S^{cc'}_{Q}\gamma_{5}Tr\Big[S^{ab'}_{Q'}S'^{ba'}_{q}\Big]
\nonumber\\&+&\beta\Big( 2\gamma_{5}S^{cb'}_{Q}\gamma_{5}S'^{aa'}_{Q'}S^{bc'}_{q}
+\gamma_{5}S^{cb'}_{Q}\gamma_{5}S'^{ba'}_{q}S^{ac'}_{Q'}- 2\gamma_{5}S^{ca'}_{Q'}
\gamma_{5}S'^{ab'}_{Q}S^{bc'}_{q}\nonumber\\&+&
\gamma_{5}S^{cb'}_{Q'}\gamma_{5}S'^{ba'}_{q}S^{ac'}_{Q}-2\gamma_{5}S^{ca'}_{q}
\gamma_{5}S'^{ab'}_{Q}S^{bc'}_{Q'}+2\gamma_{5}S^{ca'}_{q}\gamma_{5}S'^{bb'}_{Q'}
S^{ac'}_{Q}
\nonumber\\&+&2S^{cb'}_{Q}S'^{aa'}_{Q'}\gamma_{5}S^{bc'}_{q}\gamma_{5}+S^{cb'}_{Q}
S'^{ba'}_{q}\gamma_{5}S^{ac'}_{Q'}\gamma_{5}-2S^{ca'}_{Q'}S'^{ab'}_{Q}\gamma_{5}
S^{bc'}_{q}\gamma_{5}
\nonumber\\&+&S^{cb'}_{Q'}S'^{ba'}_{q}\gamma_{5}S^{ac'}_{Q}\gamma_{5}-2S^{ca'}_{q}
S'^{ab'}_{Q}\gamma_{5}S^{bc'}_{Q'}\gamma_{5}+2S^{ca'}_{q}S'^{bb'}_{Q'}\gamma_{5}
S^{ac'}_{Q}\gamma_{5}
\nonumber\\&+&4\gamma_{5}S^{cc'}_{q}Tr\Big[S^{ab'}_{Q}\gamma_{5}S'^{ba'}_{Q'}\Big]
+4S^{cc'}_{q}\gamma_{5}Tr\Big[S^{ab'}_{Q}S'^{ba'}_{Q'}\gamma_{5}\Big]
+\gamma_{5}S^{cc'}_{Q'}Tr\Big[S^{ab'}_{Q}\gamma_{5}S'^{ba'}_{q}\Big]\nonumber\\
&+&S^{cc'}_{Q'}\gamma_{5}Tr\Big[S^{ab'}_{Q}S'^{ba'}_{q}\gamma_{5}\Big]+
\gamma_{5}S^{cc'}_{Q}Tr\Big[S^{ab'}_{Q'}\gamma_{5}S'^{ba'}_{q}\Big]+S^{cc'}_{Q}
\gamma_{5}Tr\Big[S^{ab'}_{Q'}S'^{ba'}_{q}\gamma_{5}\Big]\Big)\nonumber\\&+&
\beta^2\Big( 2S^{cb'}_{Q}\gamma_{5}S'^{aa'}_{Q'}\gamma_{5}S^{bc'}_{q}+S^{cb'}_{Q}
\gamma_{5}S'^{ba'}_{q}\gamma_{5}S^{ac'}_{Q'}-
2S^{ca'}_{Q'}\gamma_{5}S'^{ab'}_{Q}\gamma_{5}S^{bc'}_{q}\nonumber\\&+&S^{cb'}_{Q'}
\gamma_{5}S'^{ba'}_{q}\gamma_{5}S^{ac'}_{Q}-
2S^{ca'}_{q}\gamma_{5}S'^{ab'}_{Q}\gamma_{5}S^{bc'}_{Q'}+2S^{ca'}_{q}\gamma_{5}
S'^{bb'}_{Q'}\gamma_{5}S^{ac'}_{Q}
\nonumber\\&+&
4S^{cc'}_{q}Tr\Big[S^{ba'}_{Q'}\gamma_{5}S'^{ab'}_{Q}\gamma_{5}\Big]+S^{cc'}_{Q'}
Tr\Big[S^{ba'}_{q}\gamma_{5}S'^{ab'}_{Q}\gamma_{5}\Big]
+
S^{cc'}_{Q}Tr\Big[S^{ba'}_{q}\gamma_{5}S'^{ab'}_{Q'}\gamma_{5}\Big]
\Big)
\Big\}\mid 0\rangle.\nonumber\\
\end{eqnarray}

For calculation of the QCD side,  we need to know the explicit expressions of the
light and heavy  quarks  propagators. Their expressions 
in coordinate space are presented in the Appendix. 

The coefficient of any structure in Eq. (\ref{str})  in QCD side can be
written as the following dispersion integral:
\begin{eqnarray}
 \Pi^{S(A)}_{1(2)}(q^2)= \int \frac{\rho^{S(A)}_{1(2)}(s)}{s-q^2} ds,
\end{eqnarray}
where $\rho^{S(A)}_{1(2)}$ are spectral densities which can be obtained
from imaginary parts of the correlation functions, i.e.,
\begin{eqnarray}
 \rho^{S(A)}_{1(2)}(s)=\frac{1}{\pi}Im\Bigg\{\Pi^{S(A)}_{1(2)}(s)\Bigg\},
\end{eqnarray}
where subindex $1(2)$ corresponds to the coefficient of the structure
$\rlap/{q}(U)$.
 
Our main task now is to calculate these spectral densities. For this aim,
we write integral representation of the Bessel functions  in Euclidean space,
\begin{eqnarray}
 \frac{K_\nu(m_Q\sqrt{-x^2})}{(\sqrt{-x^2})^\nu}=\frac{1}{2}
\int\frac{dt}{t^{\nu+1}} e^{-\frac{m_Q}{2}(t+\frac{x_E^2}{t})},
\end{eqnarray}
where $x_E^2=-x^2$. We also use the Schwinger representation for the terms containing
$\frac{1}{(x^2)^n}$ to write them as exponential form after Wick rotation, i.e.,
\begin{eqnarray}
 \frac{1}{(x_E^2)^n}=\frac{1}{\Gamma(n)}\int_0^\infty d\lambda \lambda^{n-1}
e^{-\lambda x_E^2}.
\end{eqnarray}
Then, we perform the Gaussian integrals $d^Dx_E$ and  make dimensional
regularization. After lengthy calculations for the spectral densities, we get the results presented in the Appendix.

As has already been noted, the mass sum rules for the heavy baryons are also calculated in 
\cite{Rbwtq11,Rbwtq12} and \cite{Rbwtq13}. Here we should remind that although the general form of the
interpolating current is used in \cite{Rbwtq13} similar to our case, the results
are presented only for the symmetric current case and the anti-symmetric current with
respect to the heavy quarks is not analyzed.
On the other hand, the results presented in \cite{Rbwtq11} and \cite{Rbwtq12} are obtained only for the Ioffe
current $(\beta=-1)$. Here we would like to compare our results that are obtained using the symmetric current with those presented in \cite{Rbwtq13} as well as our predictions 
in the case of
special Ioffe
current with those of for instance \cite{Rbwtq11} in details.

Comparing our results on the spectral densities for the symmetric current case with
those given in \cite{Rbwtq13}, we see that as far as the $\rlap/{q}$ structure is considered, the
perturbative parts coincide, but there is a sign difference in the terms
containing $\qq$. In the case of the structure $U$, the situation
is reversed, i.e., the signs of the perturbative parts in two works are different while the terms
containing $\qq$ have the same sign. Moreover, the terms containing the quark condensates multiplied by $m_q$
factor also show the same differences. For both structures, the coefficients of the
operator $m_0^2 \qq$  in our case are totally different from those presented in \cite{Rbwtq13}.

Comparing our results for the Ioffe case and symmetric current with the results of \cite{Rbwtq11},
we observe that the coefficient of the $m_Q m_{Q^\prime}$  factor in spectral density
$\rho^S_1(s)$ in \cite{Rbwtq11} contains an extra multiplying factor of 4. Moreover, the quark
condensate term is multiplied by the factor $5/8$, while in our case this factor is $1/2$.

Now let us compare the results for  the spectral density $\rho^S_2(s)$.
In our case, there is no any term multiplied by only the light quark mass $m_q$ (without condensates), while in
\cite{Rbwtq11} there exists such a term (first term in Eq. (26) in \cite{Rbwtq11}). When we
check the term containing the quark condensate multiplied by $m_Q m_{Q^\prime}$, our
result is half of that given in \cite{Rbwtq11}.

In the anti-symmetric current case, we obtain the following differences  
among our results compared to those given in \cite{Rbwtq11}.

For the spectral density $\rho^A_1(s)$:

\begin{itemize}
\item the coefficient of the term $m_Q m_{Q^\prime}$ in \cite{Rbwtq11} contains an extra
factor of $2$.

\item The quark condensate term in our case is multiplied by $1/4$, but it 
is multiplied by $-5/32$ in \cite{Rbwtq11}.

\item We have the term,
\bea
\label{nolabel}
{1\over 12} \qq \Big[ (1-\alpha) m_Q + \alpha m_{Q^\prime} \Big], \nnb
\eea
 in our result, which is totally absent in \cite{Rbwtq11}.

\item Also, the term
\bea
\label{nolabel} 
{1\over 16} {m_q\over \alpha \beta} (\beta m_{Q^\prime} + \alpha m_Q), \nnb
\eea
is absent in \cite{Rbwtq11}.
\end{itemize}

When the spectral density $\rho^A_2(s)$ is analyzed, we see that
\begin{itemize}

\item the term,
\bea
\label{nolabel}
{1\over 32 \pi^4} {1\over \alpha^2 \beta^2} (\alpha m_{Q^\prime} + 
\beta m_Q), \nnb
\eea
appears in our case, but is absent in \cite{Rbwtq11}.

\item The coefficients of the operators $\qq$ and $m_0^2 \qq$  in \cite{Rbwtq11} are also 
different compared to our results.

\end{itemize}

Our final task in this section is to equate the coefficients of the selected
structures from both physical and QCD sides to obtain QCD sum rules for the
mass and residue of the doubly heavy baryons under consideration.
After performing Borel transformation with respect to the variable-$q^2$ as
well as continuum subtraction  to suppress contribution of the higher states
and continuum; and using quark-hadron duality assumption we get
\begin{eqnarray}\label{sum1}
 \lambda^{2}_{B^{S(A)}} e^{\frac{-m^{2}_{B^{S(A)}}}{M^2}}&=&\int_{(m_Q+
m_{Q'})^2}^{s_0} ds \rho^{S(A)}_{1}(s)  e^{\frac{-s}{M^2}},\nnb\\
\lambda^{2}_{B^{S(A)}} m_{B^{S(A)}}e^{\frac{-m^{2}_{B^{S(A)}}}{M^2}}&=&
\int_{(m_Q+m_{Q'})^2}^{s_0} ds \rho^{S(A)}_{2}(s)  e^{\frac{-s}{M^2}},
\end{eqnarray}
where, $M$ and $s_0$ are Borel mass parameter and continuum threshold,
respectively. Obviously, the mass sum rule for the baryons under consideration
is obtained eliminating the residue in each of the above equations.
This is possible applying derivative with respect to the $-\frac{1}{M^2}$ to
both sides of each   equation in Eq. (\ref{sum1}) and dividing by themselves.
As a result, we get
\begin{eqnarray}\label{sum2}
 m^{2}_{B^{S(A)}}=\frac{\int_{(m_Q+m_{Q'})^2}^{s_0} ds s\rho^{S(A)}_{i}(s)
e^{\frac{-s}{M^2}}}{\int_{(m_Q+m_{Q'})^2}^{s_0} ds \rho^{S(A)}_{i}(s)
e^{\frac{-s}{M^2}}},
\end{eqnarray}
where, $i$ can be either $1$ or $2$.

\section{Numerical results}
In this section, we present our numerical results on the mass and residues
of the doubly heavy spin--1/2 baryons. For the heavy quark masses, we use 
their $\overline{\mbox{MS}}$ masses, which are given as, 
$\bar{m}_c(\bar{m}_c) = (1.28 \pm 0.03)~GeV$,
$\bar{m}_b(\bar{m}_b) = (4.16 \pm 0.03)~GeV$ \cite{Rbwtq15}
and $m_s(2 ~GeV) = (102\pm 8)~GeV$ \cite{Rbwtq16}. The values of the quark 
condensates are taken as $\bar uu (1~GeV) = \bar dd (1~GeV) =- (246_{-19}^{+28}~MeV)^3$
\cite{Rbwtq17}, $\langle \bar s s\rangle=0.8\langle \bar u u\rangle$ and $m_0^2=(0.8\pm0.2)~GeV^2$.
It should be noted here that in our calculations we neglect the gluon
condensate contributions, since their influence to the total result is quite
small due to the fact that we have two loops in perturbative part and
radiation of gluons from heavy quark propagators contains an extra
$\alpha_s$ and $1/m_Q$ suppression. This expectation is confirmed by
explicit calculation in \cite{Rbwtq12}, and it is shown that the gluon
contribution constitutes only about 2\% of the total result. Therefore,
gluon condensate contribution can safely be neglected. Here  we should also stress that the   two different structures under consideration give approximately the same
results, hence in the following, we present the results extracted for the  structure $\rlap/{q}$.

The sum rules for physical quantities contain also three auxiliary parameters,
namely Borel mass parameter $M^2$ continuum threshold $s_0$ and general
parameter $\beta$ entered the general spin--1/2 currents. We shall find
their working region such that the masses and residues are practically 
independent of these parameters according to the standard criteria in
QCD sum rules.

The continuum threshold $s_0$ is not totally arbitrary but depends on the
energy of the first excited state. As we have not enough information about
the first excited states of these baryons, we take the value of the continuum
threshold to be in the intervals $s_0=(108-120)~GeV^2$ for $bb$,
$s_0=(45-56)~GeV^2$ for $bc$ and $s_0=(13-21)~GeV^2$ for $cc$ baryons in
accordance with the baryons containing a single  heavy
quark. Note that there is an approach which introduces the effective
thresholds, i.e., it is assumed that the continuum threshold $s_0$ is
dependent on $Q^2$ \cite{Rbwtq18}. But in the present work we will follow the
standard procedure, i.e., $s_0$ is independent of $Q^2$.
Our numerical results show that in the presented intervals for different doubly heavy
baryons, the physical quantities under consideration weakly depend on the continuum
threshold.

The upper bound on the Borel mass parameter
$M^2$ is obtained from the condition that the pole
contribution is larger compared to continuum and higher states.
For this aim, we consider the ratio,
\bea
\label{nolabel}
R = {\ds \int_{(m_Q+m_{Q^\prime})^2}^{s_0}\ds  \rho(s) e^{-s/M^2} \over
\ds \int_{(m_Q+m_{Q^\prime})^2}^\infty \rho(s) e^{-s/M^2}}, 
\eea
which describes the relative contributions of the continuum and pole.
Demanding that $R > 1/2$ (i.e., the pole contribution exceeds the contributions of the
 higher states and continuum), we obtain the maximum values of $M^2$
for different channels as follows:

\bea
\label{e8202}
M_{max}^2 = \left\{ \begin{array}{c}
6~GeV^2~(\mbox{at}~\sqrt{s_0}=4.2~GeV),~\mbox{for}~\Xi_{cc}~\mbox{and}~\Omega_{cc} \\
9~GeV^2~(\mbox{at}~\sqrt{s_0}=7.5~GeV),~\mbox{for}~\Xi_{bc}~\mbox{and}~\Omega_{bc} \\
15~GeV^2~(\mbox{at}~\sqrt{s_0}=10.9~GeV),~\mbox{for}~\Xi_{bb}~\mbox{and}~\Omega_{bb}.
\end{array} \right.
\eea

The lower bound on $M^2$ is determined from the condition that the
perturbative contribution should be larger compared to the nonperturbative
contribution. As a result of the analysis of this restriction we get,   
\bea
\label{e8202}
M_{min}^2 = \left\{ \begin{array}{c}            
3~GeV^2,~\mbox{for}~\Xi_{cc}~\mbox{and}~\Omega_{cc} \\                 
6~GeV^2,\mbox{for}~\Xi_{bc}~\mbox{and}~\Omega_{bc} \\
10~GeV^2,~\mbox{for}~\Xi_{bb}~\mbox{and}~\Omega_{bb},               
\end{array} \right.
\eea
at aforementioned values of $\sqrt{s_0}$.

Our numerical analysis shows that in these ``working regions" of $M^2$,
the perturbative contributions also exceed the nonperturbative contributions. For
example, for the $\Xi_{bb}$ baryon, at $\sqrt{s_0}=10.9~GeV$ and at
$M^2=11~GeV^2$ the contribution form perturbative part constitutes about
70\% of the total result, while the higher states and nonperturbative
contributions constitute about 30\% of the total result. Similar results are
observed for all considered baryons.

As an example in Fig. (1) we present the dependence of the mass of
the $\Xi_{bb}$ baryon
on $M^2$ at the fixed value of $\sqrt{s_0}=10.9~GeV$ and at six different values
of the parameter $\beta$. From this figure, we see that the mass of the
$\Xi_{bb}$ baryon exhibits a good stability with respect to the variation in
$M^2$.
\begin{figure}
\vskip 1.5 cm
    \includegraphics{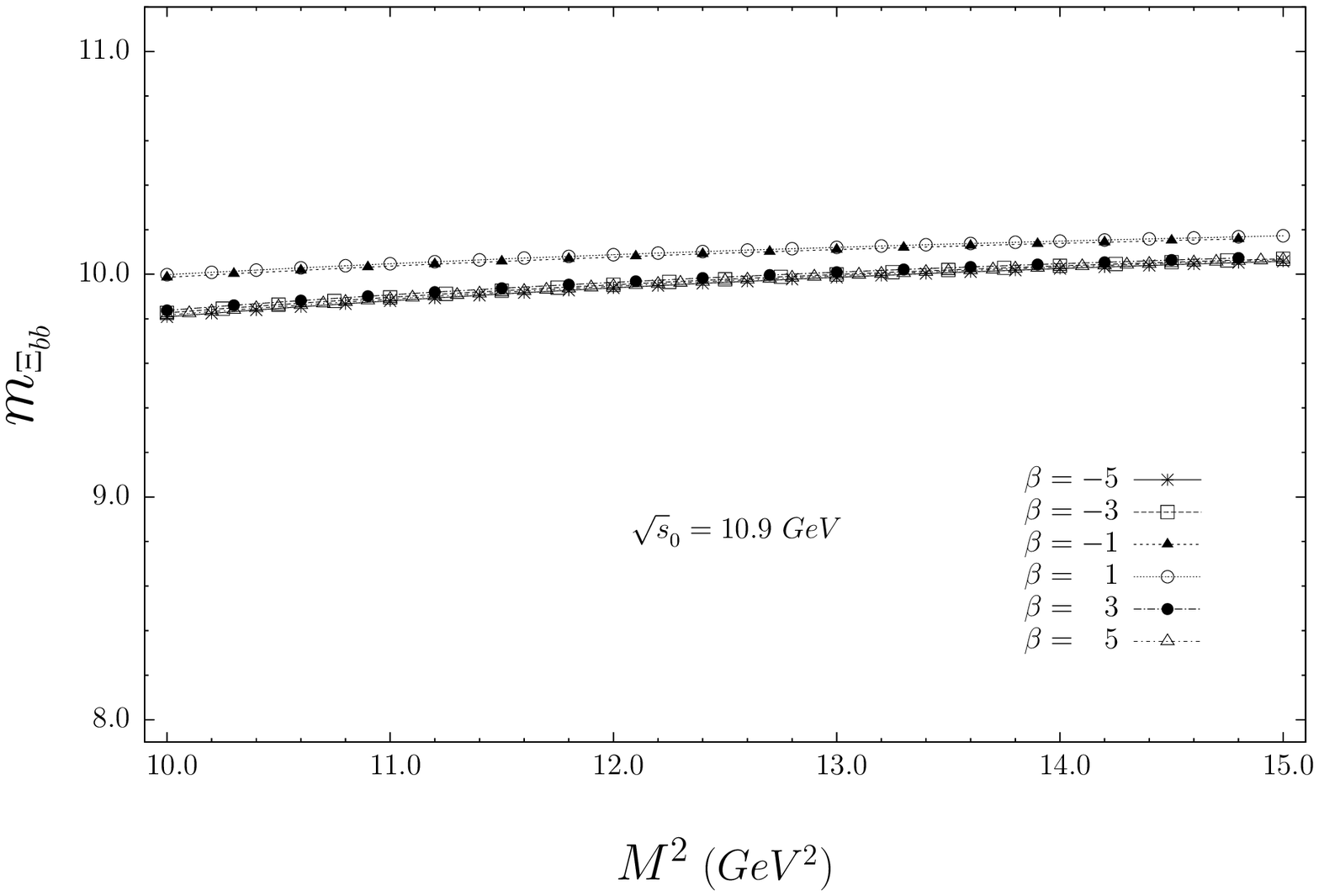}
\vskip 7.8cm
\caption{The dependence of the mass of the $\Xi_{bb}$ baryon
on $M^2$ at the fixed value of $\sqrt{s_0}=10.9~GeV$ and at six different values
of the parameter $\beta$.}
\end{figure}  

\begin{figure}   
\vskip 1.5 cm
    \includegraphics{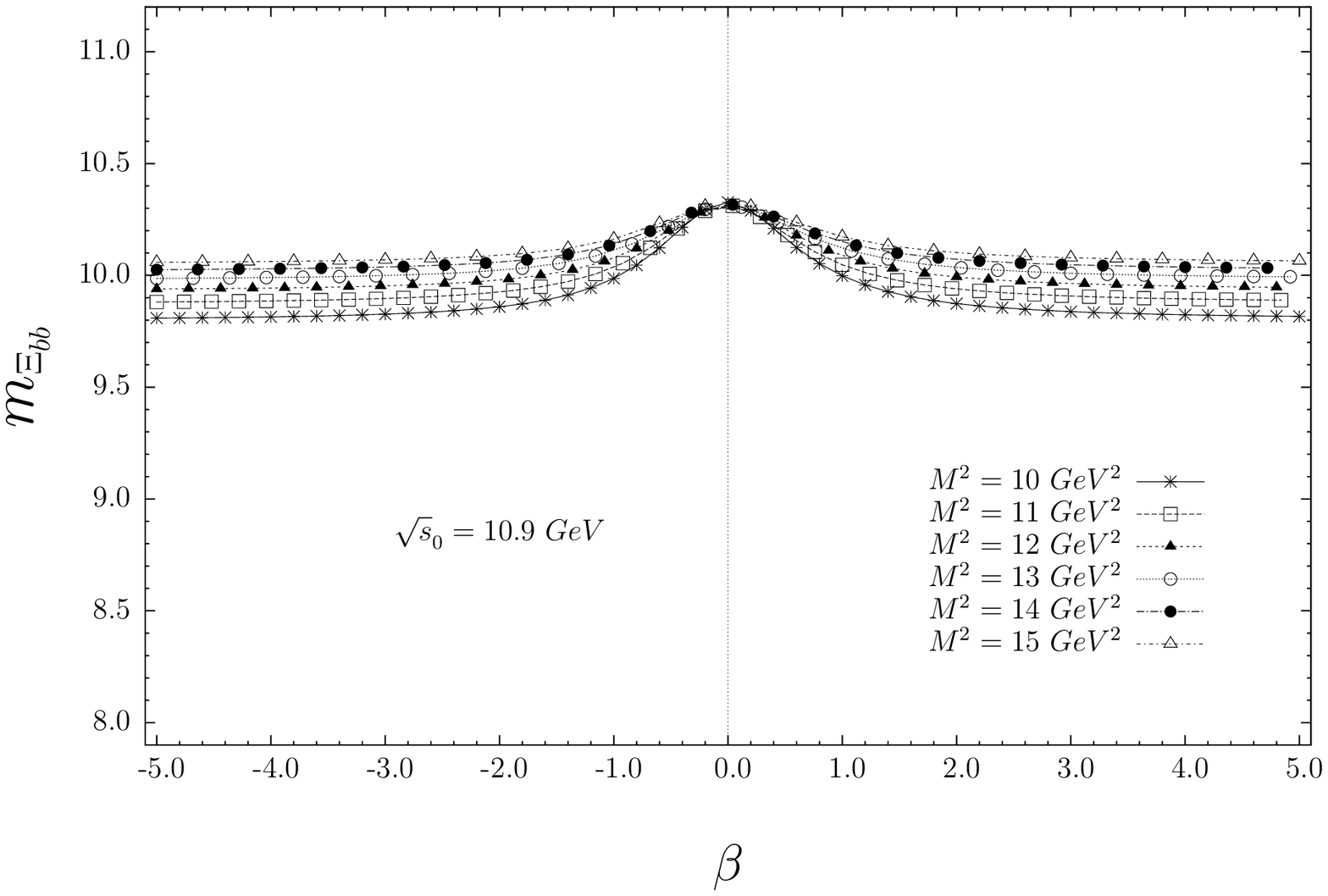}
\vskip 7.8 cm      
\caption{The dependence of the mass of the $\Xi_{bb}$ baryon
on $\beta$ at the fixed value of $\sqrt{s_0}=10.9~GeV$ and at six different
values of $M^2$.}
\end{figure}

Having determined the working region of $M^2$ and the values of
$\sqrt{s_0}$, our final attempt is to find the working region of the
arbitrary parameter $\beta$.
Depicted in Figure (2) is the dependence of the mass of the $\Xi_{bb}$
baryon on $\beta$. We see from this figure that when $\beta \ge 2$ and
$\beta \le -2$ the mass of the $\Xi_{bb}$ baryon depends very weakly on
$\beta$ and we obtain $m_{\Xi_{bb}}= 9.96 \pm 0.90~GeV$ at $M^2=11~GeV^2$
and $\sqrt{s_0}= 10.9~GeV$.

Performing similar analysis for other baryons by using the above--mentioned
regions for the auxiliary parameters, we obtain the numerical values of
their masses which are presented in Table 1.   

\begin{landscape}
\begin{table}[h]
\renewcommand{\arraystretch}{1.3}
\addtolength{\arraycolsep}{1pt}
$$
\begin{array}{|c|c|c|c|c|c|c|c|c|c|c|c|}
\hline \hline
       \mbox{Baryon}         & M^2 & \sqrt{s_0} &  \mbox{This work}  
& \mbox{\cite{Rbwtq11}}& \mbox{\cite{Rbwtq12}}&\mbox{\cite{Rbwtq13}}&\mbox{\cite{Rbwtq06}}&\mbox{\cite{Rbwtq19}}&\mbox{Exp \cite{Rbwtq20}}\\
\hline\hline
 \Xi_{bb}          &  11.0 & 10.9 & 9.96(0.90)   & 9.78(0.07)  & 10.17(0.14) & 9.94(0.91)   &  10.202   &   -   &   -            \\
 \Omega_{bb}       &  11.0 & 10.9 & 9.97(0.90)   & 9.85(0.07)  & 10.32(0.14) & 9.99(0.91)   &  10.359   &   -   &   -            \\
 \Xi_{bc}          &  8.0  &  7.5 & 6.72(0.20)   & 6.75(0.05)  &   -         & 6.86         &   6.933   &7.053  &   -            \\
 \Omega_{bc}       &  8.0  &  7.5 & 6.75(0.30)   & 7.02(0.08)  &   -         & 6.864        &   7.088   &7.148  &   -            \\
\Xi_{cc}           &  5.0  &  4.6 & 3.72(0.20)   & 4.26(0.19)  & 3.57(0.14)  & 3.52(0.06)   &   3.620   &3.676  & 3.5189(0.0009) \\
 \Omega_{cc}       &  5.0  &  4.6 & 3.73(0.20)   & 4.25(0.20)  & 3.71(0.14)  & 3.53(0.06)   &   3.778   &3.787  &   -            \\
 \Xi'_{bc}         &  8.0  &  7.5 & 6.79(0.20)   & 6.95(0.08)  &   -         &   -          &   6.963   &7.062  &   -            \\
 \Omega'_{bc}      &  8.0  &  7.5 & 6.80(0.30)   & 7.02(0.08)  &   -         &   -          &   7.116   &7.151  &   -            \\
 \hline \hline
\end{array}
$$
\caption{The mass of the doubly heavy spin--1/2 baryons (in units of $GeV$)
at $\beta = \pm 2$.} \label{tab:13}

\renewcommand{\arraystretch}{1}
\addtolength{\arraycolsep}{-1.0pt}
\end{table}
\end{landscape}

For comparison, we also present  predictions of some other theoretical papers 
as well as existing experimental data on mass of $\Xi_{cc}$ doubly charmed
baryon. The errors in the values of the present work
also belong to the uncertainties in determination of the working regions
for different  auxiliary parameters, as well as uncertainties in the values
of the input parameters.
From Table 1, We observe  that different approaches predict very close results
on the masses of the doubly heavy baryons.  Note that although there are some discrepancies between our results on the expressions of the spectral densities with those of the \cite{Rbwtq11,Rbwtq12}
 and \cite{Rbwtq13} discussed in the previous section, considering the uncertainties of the values presented in this Table, there is a good consistency between our numerical values and those existing in \cite{Rbwtq11,Rbwtq12}
 and \cite{Rbwtq13} which apply the same approach to calculate the masses of the doubly heavy baryons.

We also present the numerical values for the residues of the doubly heavy
baryons in the present work together with the existing prediction of
\cite{Rbwtq11} and \cite{Rbwtq12} in Table 2. The values presented inside the parenthesis  show the uncertainties of the results on the residues. In obtaining the values of
residues of the doubly heavy baryons, we use the values of $M^2$ and $\sqrt{s_0}$
that are presented in Table 1, at $\beta = \pm 2$.  
It follows from Table 2 that our
predictions on the residues are quite close to the results of \cite{Rbwtq12},
except the values of residues for $\Xi_{bb}$ and $\Omega_{bb}$. But our predictions and
the predictions of \cite{Rbwtq12} on the residues, are considerably different
compared to the ones given in \cite{Rbwtq11}.

\begin{table}[h]
\renewcommand{\arraystretch}{1.0}
\addtolength{\arraycolsep}{1pt}
$$
\begin{array}{|c|c|c|c|}

\hline \hline
\mbox{Baryon}      & \mbox{Present work}  & \mbox{\cite{Rbwtq11}}  & \mbox{\cite{Rbwtq12}}\\
\hline\hline
 \Xi_{bb}          & 0.44(0.08)           & 0.067 \div 0.057   & 0.252(0.064)     \\
 \Omega_{bb}       & 0.45(0.08)           & -                  & 0.311(0.077)     \\
 \Xi_{bc}          & 0.28(0.05)           & 0.046 \div 0.021   & -                \\
 \Omega_{bc}       & 0.29(0.05)           & -                  & -                \\
\Xi_{cc}           & 0.16(0.03)           & 0.042 \div 0.026   & 0.115(0.027)     \\
 \Omega_{cc}       & 0.18(0.04)           & -                  & 0.138(0.030)     \\
 \Xi'_{bc}         & 0.30(0.05)           & -                  & -                \\
 \Omega'_{bc}      & 0.31(0.06)           & -                  & -                \\
 \hline \hline
\end{array}
$$
\caption{The residues of the doubly heavy spin--1/2 baryons (in units of
$GeV^3$).} \label{tab:131}
\renewcommand{\arraystretch}{1}
\addtolength{\arraycolsep}{-1.0pt}
\end{table}

At the end of this section we would like to mention that according to Eq. (\ref{sum1}), there is another way for determination of the masses of the heavy baryon, i.e.,
\begin{eqnarray}\label{sum2}
 m_{B^{S(A)}}=\frac{\int_{(m_Q+m_{Q'})^2}^{s_0} ds \rho^{S(A)}_{2}(s)
e^{\frac{-s}{M^2}}}{\int_{(m_Q+m_{Q'})^2}^{s_0} ds \rho^{S(A)}_{1}(s)
e^{\frac{-s}{M^2}}}.
\end{eqnarray}
Our numerical calculations show that, using this formula, the central values of the masses presented in Table 1 are changed  maximally 5\%. 

In summary, we have calculated the masses and residues of the doubly heavy spin--1/2 baryons considering both symmetric and anti-symmetric currents with respect to the exchange of heavy quarks  in their most general forms in the framework of QCD sum rules.
Our results are in a good consistency with the predictions of the other theoretical papers as well as the experimental data on the mass of the doubly-charmed  $\Xi_{cc}$ baryon. We hope that the LHC
will provide possibility to study the doubly heavy baryons  in near future.
\newpage

\section{Appendix}

In this appendix, we present the expressions of the light and heavy quarks propagators as well as the spectral densities.

In our calculations, we have used the following expression for the  light quark propagator:
\bea
\label{eh32v18}
S_q(x) \es {i \rlap/x\over 2\pi^2 x^4} - {m_q\over 4 \pi^2 x^2} -
{\lla \bar q q \rra\over 12} \left(1 - i {m_q\over 4} \rlap/x \right) -
{x^2\over 192} m_0^2 \lla \bar q q \rra  \left( 1 -
i {m_q\over 6}\rlap/x \right) ~.
\eea
 The heavy quark propagator  in an external  field is given as 
\bea
\label{eh32v19}
S_Q(x) &=& {m_Q^2 \over 4 \pi^2} {K_1(m_Q\sqrt{-x^2}) \over \sqrt{-x^2}} -
i {m_Q^2 \rlap/{x} \over 4 \pi^2 x^2} K_2(m_Q\sqrt{-x^2})\nnb \\& -&
ig_s \int {d^4k \over (2\pi)^4} e^{-ikx} \int_0^1
du \Bigg[ {\rlap/k+m_Q \over 2 (m_Q^2-k^2)^2} G^{\mu\nu} (ux)
\sigma_{\mu\nu} +
{u \over m_Q^2-k^2} x_\mu G^{\mu\nu} \gamma_\nu \Bigg]~,\nnb \\
\eea
where $K_1$ and $K_2$ are the modified Bessel function of the second kind. 

The explicit expressions for the  spectral densities obtained after procedure mentioned in the body text are given as:
\begin{eqnarray}
\rho^{S}_{1}(s)&=&\frac{A}{128 \pi^4}\int_{\psi_{min}}^{\psi_{max}}
\int_{\eta_{min}}^{\eta_{max}}d\psi d\eta\Bigg\{3 \mu \Bigg[\psi\eta
\Big[5+\beta (2+5\beta)\Big]\mu+2 (-1+\psi+\eta)(-1+\beta)^2 m_Q m_{Q'}\nnb\\
&-&6(-1+\beta^2)m_q(\eta m_Q+\psi m_{Q'})\Bigg]\Bigg\}+
\frac{A\langle \bar q q\rangle}{16 \pi^2}\int_{\psi_{min}}^{\psi_{max}}d\psi
\Bigg\{-\Bigg[(-1+\psi)\psi\Big[5+\beta(2+5\beta)\Big]m_q\Bigg]\nnb\\
&+&3(-1+\beta^2)\Big[(-1+\psi)m_Q-\psi m_{Q'}\Big]\Bigg\},
\end{eqnarray}

\begin{eqnarray}
 \rho^{A}_{1}(s)&=&\frac{1}{256 \pi^4}\int_{\psi_{min}}^{\psi_{max}}
\int_{\eta_{min}}^{\eta_{max}}d\psi d\eta\Bigg\{ \mu \Bigg[3\psi\eta
\Big[5+\beta (2+5\beta)\Big]\mu+2 (-1+\beta)\Big[(-1+\psi+\eta)
(13\nnb\\&+&11\beta) m_Q m_{Q'}
-(1+5\beta)m_q(\eta m_Q+\psi m_{Q'})\Big]\Bigg]\Bigg\}\nnb\\
&+&\frac{\langle \bar q q\rangle}{96 \pi^2}\int_{\psi_{min}}^{\psi_{max}}
d\psi\Bigg\{-3(-1+\psi)\psi\Big[5+\beta(2+5\beta)\Big]m_q\nnb\\
&+&(-1+\beta)(1+5\beta)\Big[(-1+\psi)m_Q-\psi m_{Q'}\Big]\Bigg\},
\end{eqnarray}
\begin{eqnarray}
 \rho^{S}_{2}(s)&=&\frac{A}{128 \pi^4}\int_{\psi_{min}}^{\psi_{max}}
\int_{\eta_{min}}^{\eta_{max}}d\psi d\eta\Bigg\{3 \mu 
\Bigg[3\psi(-1+\beta^2)\mu m_{Q'}+m_Q\Big[ 3\eta(-1+\beta^2) \mu\nnb\\
&-&2\Big[5+\beta(2+5\beta)\Big]m_q m_{Q'}\Big]\Bigg]\Bigg\}+
\frac{A\langle \bar q q\rangle}{32 \pi^2}\int_{\psi_{min}}^{\psi_{max}}
d\psi\Bigg\{-\Bigg[(-1+\psi)\psi(-1+\beta)^2\Big[3m_0^2\nnb\\
&+&4\mu'-2 s\Big]+2\Big[5+\beta(2+5\beta)\Big]m_Q m_{Q'}+6(-1+\beta^2)
m_q\Big[(-1+\psi)m_Q-\psi m_{Q'}\Big]\Bigg]\nnb\\&-&\frac{3}{4}m_0^2(1-\beta)^2\Bigg\},
\end{eqnarray}
\begin{eqnarray}
 \rho^{A}_{2}(s)&=&\frac{1}{256 \pi^4}\int_{\psi_{min}}^{\psi_{max}}
\int_{\eta_{min}}^{\eta_{max}}d\psi d\eta\Bigg\{ \mu \Bigg[\psi(-1+\beta)
(1+5\beta)\mu m_{Q'}+ m_{Q}\Big[\eta(-1+\beta)(1+5\beta)\mu
\nnb\\&-&6[5+\beta(2+5\beta)]m_q m_{Q'}\Big]\Bigg]\Bigg\}+\frac{\langle
\bar q q\rangle}{192 \pi^2}\int_{\psi_{min}}^{\psi_{max}}d\psi
\Bigg\{-\Bigg[(-1+\psi)\psi(-1+\beta)(13+11\beta)\Big[3m_0^2\nnb\\
&+&4\mu'-2 s\Big]+6\Big[5+\beta(2+5\beta)\Big]m_Q m_{Q'}+2(-1+\beta)
(1+5\beta)m_q\Big[(-1+\psi)m_Q-\psi m_{Q'}\Big]\Bigg]\nnb\\&+&\frac{3}{2}m_0^2(1-\beta)^2\Bigg\},
\end{eqnarray}
where,
\begin{eqnarray}
 \mu&=&\frac{m_Q^2}{\psi}+\frac{m_{Q'}^2}{\eta}-s,\nnb\\
\mu'&=&\frac{m_Q^2}{\psi}+\frac{m_{Q'}^2}{1-\psi}-s,\nnb\\
\eta_{min}&=&\frac{\psi m_{Q'}^2}{s\psi-m_Q^2},\nnb\\
\eta_{max}&=&1-\psi,\nnb\\
\psi_{min}&=&\frac{1}{2s}\Big[s+m_Q^2-m_{Q'}^2-\sqrt{(s+m_Q^2-m_{Q'}^2)^2-
4m_Q^2s}\Big],\nnb\\
\psi_{max}&=&\frac{1}{2s}\Big[s+m_Q^2-m_{Q'}^2+\sqrt{(s+m_Q^2-m_{Q'}^2)^2-
4m_Q^2s}\Big].
\end{eqnarray}

\end{document}